\documentclass[reprint,longbibliography,superscriptaddress,amsmath,amssymb,aps,prx]{revtex4-2}

\usepackage{graphicx}
\usepackage{subfigure}
\usepackage{dcolumn}
\usepackage{bm}
\usepackage{physics}
\usepackage{color}
\usepackage{float}
\usepackage{longtable}
\usepackage{threeparttablex} 
\usepackage{booktabs} 

\usepackage{xcolor}
\DeclareRobustCommand{\erase}{\bgroup\markoverwith{\textcolor{red}{\rule[.5ex]{2pt}{0.4pt}}}\ULon}

\begin{document}
\title{Logical entanglement distribution between distant 2D array qubits}

\author{Yuya Maeda}
\affiliation{Graduate School of Engineering Science, Osaka University, 1-3 Machikaneyama, Toyonaka, Osaka 560-8531, Japan}

\author{Yasunari Suzuki}
\affiliation{NTT Computer and Data Science Laboratories, NTT Corporation, Musashino 180-8585, Japan}
\affiliation{JST, PRESTO, 4-1-8 Honcho, Kawaguchi, Saitama, 332-0012, Japan}
\affiliation{Center for Quantum Computing, RIKEN, Wako Saitama 351-0198, Japan}

\author{Toshiki Kobayashi}
\affiliation{Graduate School of Engineering Science, Osaka University, 1-3 Machikaneyama, Toyonaka, Osaka 560-8531, Japan}
\affiliation{Center for Quantum Information and Quantum Biology, Osaka University, Japan}

\author{Takashi Yamamoto}
\affiliation{Graduate School of Engineering Science, Osaka University, 1-3 Machikaneyama, Toyonaka, Osaka 560-8531, Japan}
\affiliation{Center for Quantum Information and Quantum Biology, Osaka University, Japan}

\author{Yuuki Tokunaga}
\affiliation{NTT Computer and Data Science Laboratories, NTT Corporation, Musashino 180-8585, Japan}
\affiliation{Faculty of Information Science and Technology, Hokkaido University,
Sapporo 060-0814, Japan}

\author{Keisuke Fujii}
\affiliation{Graduate School of Engineering Science, Osaka University, 1-3 Machikaneyama, Toyonaka, Osaka 560-8531, Japan}
\affiliation{Center for Quantum Information and Quantum Biology, Osaka University, Japan}
\affiliation{Center for Quantum Computing, RIKEN, Wako Saitama 351-0198, Japan}

\date{\today}

\begin{abstract}
Sharing logical entangled pairs between distant quantum nodes is a key process to achieve fault-tolerant quantum computation and communication. However, there is a gap between current experimental specifications and theoretical requirements for sharing logical entangled states while improving experimental techniques.
Here, we propose an efficient logical entanglement distribution protocol based on surface codes for two distant 2D qubit array with nearest-neighbor interaction. 
A notable feature of our protocol is that it allows post-selection according to error estimations, which provides the tunability between the infidelity of logical entanglements and the success probability of the protocol.
With this feature, the fidelity of encoded logical entangled states can be improved by sacrificing success rates.
We numerically evaluated the performance of our protocol and the trade-off relationship, and found that our protocol enables us to prepare logical entangled states while improving fidelity in feasible experimental parameters.
We also discuss a possible physical implementation using neutral atom arrays to show the feasibility of our protocol.
\end{abstract}

\maketitle

\section{introduction}
\label{sec:introduction}
Thanks to recent advances in quantum computation, quantum error correction~(QEC) with surface codes has been demonstrated with several quantum devices implemented on two-dimensional~(2D) sites, such as superconducting circuits~\cite{google_surfacecode} and neutral atoms~\cite{lukin_quantum_processor,48logical_qubits_Lukin}.
Technologies have also been developed to enable quantum communication between two distant nodes with 2D devices. Although the entanglement generation rate between a pair of physical qubits is technically limited, effective generation rates can be significantly improved by parallelizing communications in a 2D plane by optical tweezer technologies for neutral atoms~\cite{rempe_cavity_array} or the three-dimensional integration of superconducting chips~\cite{3D_integrated_qubits,entanglement_across_separate_sillicon_dies}.
Thus, our next milestone is to connect fault-tolerant quantum computers with 2D devices by combining fast communication and fault-tolerant local computation, and to ensure further scalability for fault-tolerant quantum computing~\cite{review_distributed_computation,Surface_code_quantum_communication} and quantum communication protocols~\cite{blind_quantum_computation,secret_quantum_computation}.

\begin{figure*}[htbp]
    \centering
    \includegraphics[width=\linewidth]{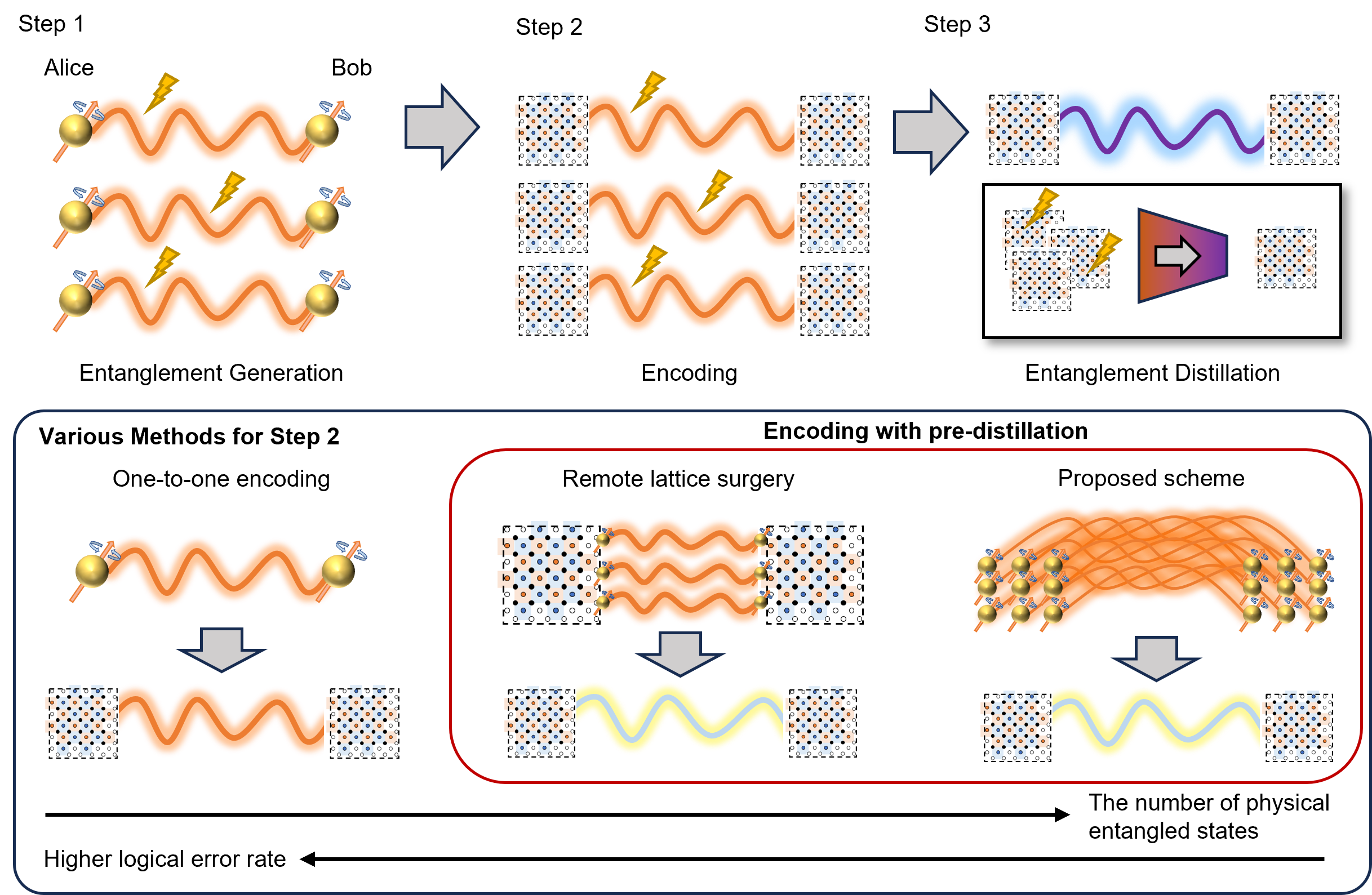}
    \caption{Three steps to generate high-fidelity logical entanglements.}
    \label{fig:sketch_of_distribution}
\end{figure*}

One of the most vital steps towards this milestone is establishing a fast and high-fidelity {\it logical entanglement distribution}~\cite{feasibility_of_logical_bell_state_generation,fault_tolerant_optical_interconnects}, which is a protocol to create an entangled state of logical qubits with a required fidelity using noisy quantum channels.
The overall protocol of the logical entanglement distribution is performed with the following three steps as shown in Fig.\,\ref{fig:sketch_of_distribution}: (i) Two nodes share several noisy physical entangled states, (ii) each node encodes shared physical qubits into logical qubits to obtain several noisy logical entangled states, and (iii) clean logical entangled states with the target fidelity are distilled from several noisy logical entangled states. It should be noted that we use the term \textit{encoding} to describe the mapping of physical entanglement into the logical code space via syndrome measurements, rather than the construction of logical states through unitary circuits.
We aim to propose a concrete protocol of the logical entanglement distribution based on established technologies and to maximize the rate of logical entanglement with a target fidelity.

In this paper, we focus on the optimization of step (ii) in the above protocol. The simplest way to achieve the encoding process in step (ii) is to encode one physical qubit into a logical qubit with a method used in magic-state preparation~\cite{li2015magic}. This strategy is used in Ref.\,\cite{pattison2025constant}, for example. The advantage of this protocol is that an arbitrarily slow physical entanglement distribution rate is acceptable since the lifetime of qubits can be extended to a sufficiently long time by quantum error correction. This is particularly useful when the lifetime of physical qubits is shorter than the period of entanglement generation.
On the other hand, if physical entangled pairs are generated faster than the consumption speed in logical entanglement distillation, this protocol cannot leverage fast physical entanglement distribution technologies. Also, the fidelity of noisy logical entangled states after step (ii) is smaller than the physical entangled states, which increases the latency and the number of required entangled states in step (iii). 

Remote lattice surgery for topological stabilizer codes~\cite{fault_tolerant_optical_interconnects,Surface_code_quantum_communication} is another candidate that can leverage fast and parallelized quantum communication. This protocol requires the edges of a 2D qubit plane to be connected by quantum channels. Supposing that logical qubits are encoded with code distance $d$, this protocol takes $d$ rounds of stabilizer measurements to detect measurement errors, and at least $O(d)$ entangled pairs are required for each round. 
Thanks to the property of quantum error correction, this protocol can generate logical entangled pairs with higher fidelity than the physical Bell pairs if the error rates of physical Bell pairs are smaller than a threshold value. The drawback of this approach is that it takes longer, i.e., $d$ round stabilizer measurement, than the baseline method. Also, this approach cannot leverage the advantage of parallelism more than a 1D array of quantum channels. Thus, more efficient logical entanglement distribution that can maximize the potential of 2D quantum devices is still lacking. While the full integration of a 2D channel array remains a significant engineering challenge, its fundamental constituents have seen remarkable progress. Notably, technologies aimed at spatially multiplexed entanglement generation, such as the simultaneous detection of photons from multiple individual atoms in an array~\cite{li2025parallelized,Shaw2026,maeda2025waveguide},have been reported, providing a solid foundation for the feasibility of our protocol in the near future.

In this paper, we propose an efficient protocol for logical quantum entanglement distribution when quantum devices are embedded in a 2D plane and each qubit is connected by a 2D quantum channel array. Our protocol consumes $O(d^2)$ physical entangled pairs to generate one logical entangled state. To handle the realistic situation of probabilistic and noisy entanglement generation, our protocol uses an adaptive post-selection according to the number of estimated errors. This post-selection makes protocol two-way but provides the tunability between the protocol success probability and the fidelity of the logical entangled states~\cite{stabilizer_distillation, distillation_and_correction}. Thus, we can choose the fast protocol with moderately distilled logical entanglement generation or the slow protocol with high-fidelity logical entanglement generation, depending on the situations and applications. If we do not postselect and accept all the events, the protocol becomes a fast one-way protocol at the cost of degraded final fidelities. This tunability is useful for tailoring the protocol to the post-distillation in step (iii) and the final target logical fidelity. We compared our protocol with the existing ones based on surface code in Table.~\ref{tab:comparing_protocols}. In this table, we list the resources required for generating one logical Bell pair.

We numerically investigated the performance of our protocol and revealed the trade-off relationship between the success rates and the fidelity of logical entanglement under several realistic parameter sets in neutral atom systems.
Our results indicate that our protocol can generate logical entangled pairs with higher fidelity than the physical entangled pairs with the current achievable experimental technologies. For example, if SWAP gate fidelity is $0.99$, error rates of logical entangled states can be less than $6.5\times10^{-3}$ with a success probability larger than $0.88$. Since our protocol is expected to be combined with the post-distillation to achieve target fidelity, we also calculated the logical-entanglement bandwidth under assumptions that would be feasible in the near future. When we target logical error rate $10^{-13}$~\cite{target_logical_fidelity}, the overall bandwidth of logical entanglement distribution is $44~{\rm Hz}$.

These results are immediately useful for designing devices, channels, and architectures of distributed quantum computing since our results provide a baseline bandwidth of practical logical entanglement distribution. Note that while we focus on fault-tolerant quantum computers with the surface codes, our theoretical framework can be straightforwardly extended to various stabilizer codes, such as good low-density parity check codes~(LDPC)~\cite{atom_qldpc,high-threshold_and_low-overhead,qLDPCcode}. Thus, our protocol will be compatible with future improvements in quantum error-correcting codes.

The contributions of our work are listed as follows.
\begin{itemize}
\item We proposed a concrete protocol to obtain logical entangled pairs that have higher fidelity than the provided physical entangled pairs and can be implemented with practical 2D devices with a 2D channel array. Using the tunable post-selection conditions, the proposed method offers tunability between the success probability and logical error rate after encoding.
\item We showed how the tunability in the encoding step should be optimized to maximize the generation rate of the target logical entangled states of the whole protocol shown in Fig.\,\ref{fig:sketch_of_distribution}, which enables us to characterize the practical evaluation of the total logical communication bandwidth with logical entangled states.
\item We numerically evaluated and optimized its performance under realistic parameter settings using neutral atoms as an example platform. Our result indicates that the achievable rate of entanglement distribution is approximately $44~{\rm Hz}$ to attain a $10^{-13}$ logical error rate.
\end{itemize}
A detailed comparison with related studies is provided in Sec.~\ref{sec:related_work} to contextualize these contributions.

This paper is organized as follows. In Sec.\,\ref{sec:protocol}, we explain our protocol to generate logical entangled pairs between two distant nodes. Then, we show the numerical results and their settings and assumptions in Sec.\,\ref{sec:evaluation}. In Sec.\,\ref{sec:discussion}, we discuss the experimental feasibility of our protocol based on the experimental reports and evaluate the logical entanglement generation rate combined with post-distillation.
In Sec.\,\ref{sec:related_work}, we discuss the positioning of our protocol within the context of related studies and clarify our unique contributions to the development of distributed quantum computation.
Finally, Sec.\,\ref{sec:conclusion} summarizes this paper and mentions future work.

\begin{table*}[t]
    \centering
    \begin{threeparttable}
        \caption{Comparison of logical entanglement generation methods based on surface code on 2D qubit-array}
        \label{tab:comparing_protocols}
        \begin{tabular}{lccccc}
            \toprule
             & Number of Bell states\tnote{a} {\hspace{4em}}& Rounds of measurement\tnote{b}{\hspace{4em}}& Pre-distillation\tnote{c}{\hspace{4em}}& Channels{\hspace{4em}}& Post-selection{\hspace{2em}}\\ 
            \midrule
            One-to-one~\cite{li2015magic} & $1$ & $1$ & No & Single channel & not available \\
            Remote surgery~\cite{fault_tolerant_optical_interconnects} & $O(d^2)$ & $d$ & Yes & 1D channel array & not considered\tnote{d} \\
            Our proposal & $O(d^2)$ & $1$ & Yes & 2D channel array & available \\
            \bottomrule
        \end{tabular}

        \begin{tablenotes}
            \small
            \item[a]  Scaling is shown relative to the code distance $d$, assuming other hardware parameters are comparable.
            \item[b]  The number of required stabilizer measurement cycles to complete the protocols
            \item[c]  Whether we can reduce error rates compared to that of physical Bell states at the encoding stage at the cost of post-selection.  
            \item[d]  Post-selection might be available in principle for remote lattice surgery, but to our knowledge, previous work implements protocols as one-way protocols, i.e., without post-selection.
        \end{tablenotes}
    
    \end{threeparttable}
\end{table*}
\section{Protocol}
\label{sec:protocol}
\begin{figure}[htbp]
    \centering
    \includegraphics[width = 0.4\textwidth]{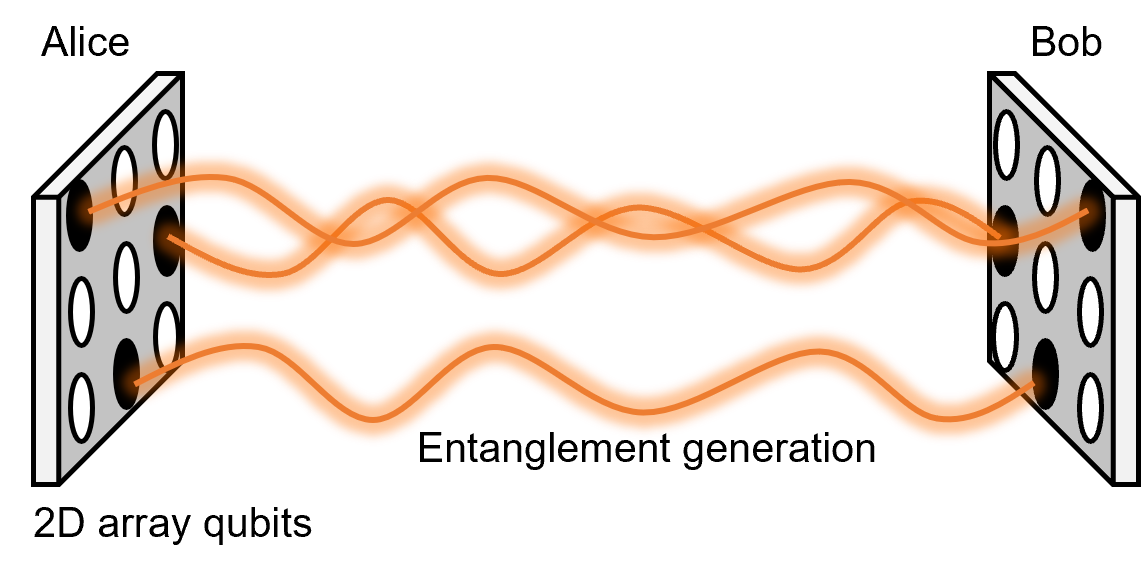}
    \caption{{\bf Entanglement generation} Alice and Bob perform multiplexed physical entanglement generation protocols between all the pairs of physical qubits at the same coordinate on the 2D square lattice. Since the protocol succeeds probabilistically, only a part of qubit pairs are successfully entangled.}
    \label{fig:sharing_entanglements}
    \includegraphics[width = 0.4\textwidth]{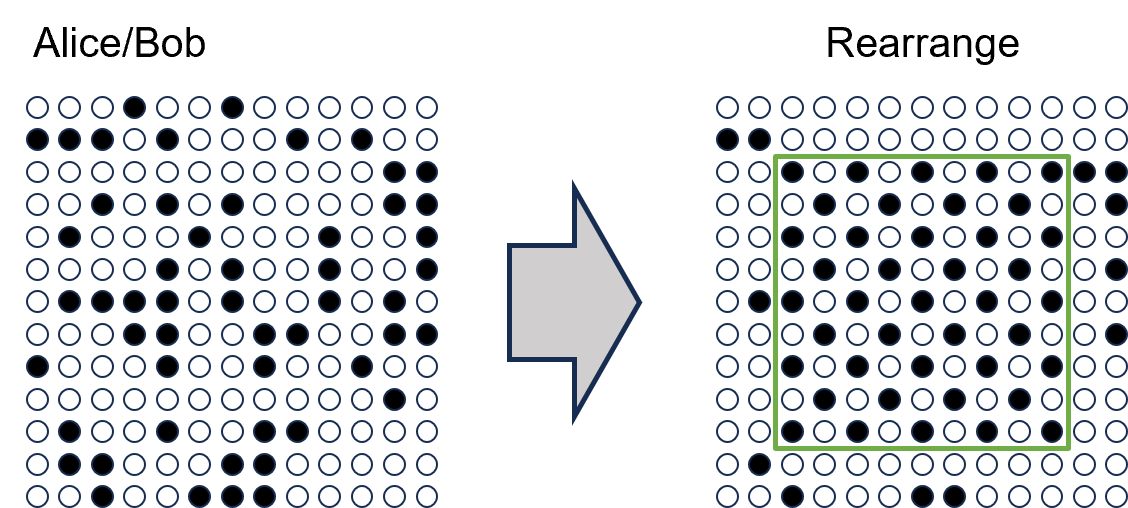}
    \caption{{\bf Qubit rearrangement} Alice and Bob rearrange the locations of generated entangled pairs so that they constitute a 2D lattice. Note that the rearrangement processes at the two nodes are the same since entangled pairs are located in the same positions.}
    \label{fig:rearrangement}
    \includegraphics[width = 0.4\textwidth]{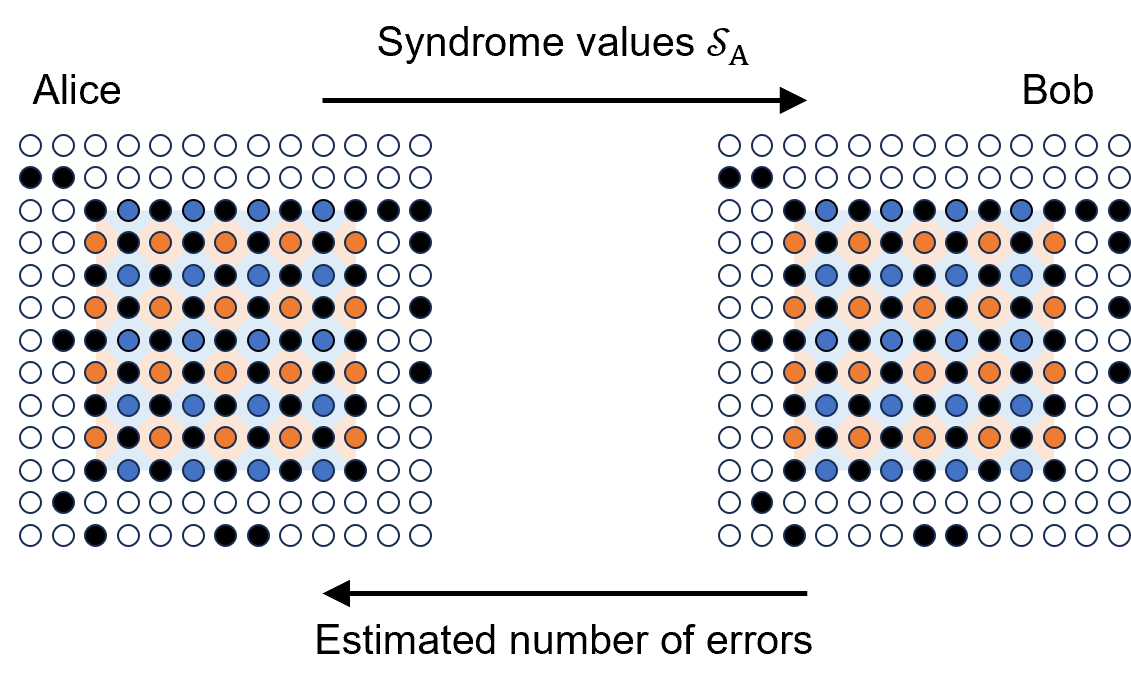}
    \caption{{\bf Syndrome measurement} Alice and Bob perform the syndrome measurements on the rearranged 2D lattice. Alice sends the obtained syndrome values to Bob, and Bob corrects errors according to the difference of syndrome values. This process projects the state of rearranged entangled pairs into the logical space of surface codes, which results in a logical entangled pair. Bob can abort and restart the protocols according to the number of estimated errors to ensure the quality of entanglement.}
    \label{fig:syndrome_measurement}
\end{figure}

In this section, we show a protocol to generate logical entangled pairs encoded with surface codes with 2D noisy quantum channels. We suppose two nodes, Alice and Bob, have 2D qubit arrays of the same size and can perform two-qubit operations on the nearest neighboring pairs of qubits. Our protocol consists of the following three steps as shown in Figs.\,\ref{fig:sharing_entanglements}, \ref{fig:rearrangement}, and \ref{fig:syndrome_measurement}.

In the following subsections, we explain the details of each step while introducing the parameters of our protocols. Throughout this paper, we use the notations listed in Table.\,\ref{tab:notation}.

\begin{table}[htbp]
    \centering
    \caption{Notations and parameters of proposed protocols.}
    \begin{tabular}{l|l}
        \hline
        $L$ & The size of 2D qubit array\\
        $p_{\rm gen}$ & Success probability of entanglement generation \\
        $e_{\rm init}$ & Error rate of physical entanglement \\
        $e_{\rm swap}$ & Error rate of local SWAP operations \\
        $w_{\rm thr}$ & Threshold of estimated errors \\
        \hline
        $p_{\rm log}$ & Success probability of our protocol \\
        $e_{\rm log}$ & Error rate of logical entanglement \\
        \hline
    \end{tabular}
    \label{tab:notation}
\end{table}

\subsection{Entanglement generation}
Suppose Alice and Bob have physical qubits aligned on the $L \times L$ 2D square lattice.
They probabilistically generate physical Bell pairs $(\ket{00}+\ket{11})/\sqrt{2}$ between two qubits at the same coordinate in parallel. 
After the generation process, the entangled qubit pairs at each node are placed on the random sites in the lattice, but the arrangements of entangled pairs on two nodes are always the same since we generate entanglement for qubit pairs at the same position. An example is shown in Fig.\,\ref{fig:sharing_entanglements}, where black pairs represent successful entangled pairs and white pairs are failed ones.

We repeat the entanglement generation for a certain duration. Then, each physical entanglement generation will succeed with finite probability $p_\mathrm{gen}$. We denote the error rate of generated entangled states as $e_\mathrm{init}$. Here, $p_{\rm gen}$ and $e_{\rm init}$ are in the trade-off relations according to the duration; the success probability $p_{\rm gen}$ increases as the number of trials increases, while the error rates $e_{\rm init}$ increases due to the decoherence.

\subsection{Qubit rearrangement}
In the rearrangement step, we determine the code distance and the position of the surface-code cell on the  2D array and then move the generated entangled qubits to the positions of the data qubits, as shown in Fig.\,\ref{fig:rearrangement}. The rearrangement schedule is described as follows.

The code distance is chosen to consume as many entangled pairs as possible. For simplicity, we assume unrotated surface codes with $[[d^2+(d-1)^2, 1, d]]$ in this paper. Let $M$ be the number of generated entangled states in the entanglement generation step. Then, the code distance is the maximum $d$ that satisfies $d^2+(d-1)^2 \le M$.

The position of the surface-code cell and the rearrangement schedule are determined so that the number of SWAP gates for moving qubits is minimized. This is because the repetitive applications of SWAP gates decrease the fidelity of entangled pairs and degrade the overall performance. To achieve this, we use the following heuristic approaches. We enumerate the first and second nearest positions from the mean position of the entangled pairs as the candidates of the center of the surface-code cell. Then, we calculate the number of SWAP gates in the rearrangement schedule for each candidate and choose the candidate with the smallest number of SWAP gates.

The schedule of qubit movement is heuristically minimized as follows. We consider a bipartite graph where the left nodes correspond to each data qubit of a surface code and the right ones to entangled qubits in Alice's (or Bob's) qubit array.
The weights between nodes are set as a distance from the data qubit to the entangled qubit. We find a minimum-weight maximum matching of this graph, and entangled qubits are moved to the matched data-qubit position.
Note that while Alice and Bob calculate the rearrangement process independently, they result in the same rearrangement process since they have the same entangled qubit positions.

In our numerical evaluation, we choose the squared Manhattan distance as the distance between the data qubit and the entangled qubit. We chose this squared weight instead of simple Manhattan or Euclid distance because Manhattan or Euclid distance sometimes generates paths with many intersections, which degrades the quality of solutions. We believe there would be better heuristic approaches for this process, but we left this as future work.

In this study, we evaluate the performance using a SWAP-gate-based rearrangement model as a versatile baseline. On the other hand, our protocol is also compatible with alternative methods such as atom shuttling via moving optical tweezers. A discussion on the potential impact of hardware-specific rearrangement techniques and their noise characteristics is provided in Sec.~\ref{sec:conclusion}.

\subsection{Syndrome measurement}
In this section, we describe the process of mapping physical Bell pairs into the logical code space. It is important to note that throughout this manuscript, we use the term \textit{encoding} to refer to this mapping procedure via syndrome measurements, rather than the construction of logical states through unitary encoding circuits.

Alice and Bob perform stabilizer Pauli measurements of surface codes to generate a logical entangled state. If there is no error, Alice and Bob will obtain the same random values, and the logical states are projected into logical Bell pairs $(\ket{0_L0_L}+\ket{1_L 1_L})/\sqrt{2}$. In practice, there are errors in a finite probability, and occurred Pauli errors must be estimated. This can be efficiently estimated from the difference in syndrome values observed in Alice's and Bob's syndrome nodes. If the errors are correctly estimated up to the stabilizer operators, the application of estimated Pauli errors will recover the logical entanglement. Thus, Alice sends the obtained syndrome values to Bob. Then, Bob estimates the physical Pauli errors and applies them to Bob's qubits. For the correctness of this procedure, please refer to Refs.\,\cite{stabilizer_distillation, distillation_and_correction}.
These studies demonstrate a general strategy: by utilizing error-correcting codes for error detection, they can be effectively applied to post-selection-based entanglement distillation. Our protocol leverages this concept to provide tunability between success probability and the resulting logical fidelity.

Before finishing the protocol, Bob judges whether the protocol is reliably finished or not. If the number of qubits affected by Pauli errors is more than an acceptance threshold $w_\mathrm{thr}$, Bob aborts and restarts the protocol and notifies the decision to Alice.
We denote the infidelity of the resultant logical entangled pair as $e_{\rm log}$. Since Bob may abort the protocol when many errors are detected, this protocol succeeds with a finite probability denoted by $p_{\rm log}$.

The acceptance threshold $w_\mathrm{thr}$ plays a role in tuning the balance between $e_{\rm log}$ and $p_{\rm log}$. If we set $w_\mathrm{thr} = 0$, the protocol fails only when there are no less than $d$ errors, which means $e_{\rm log}$ becomes small with sacrificing $p_{\rm log}$. The opposite case is choosing $w_\mathrm{thr}$ equal to the number of data qubits. Then, this protocol is never aborted, i.e., $p_{\rm log} = 1$, but the resultant logical error rates will be degraded. Note that, in this opposite case, Bob does not need to send the decision to Alice, and the protocol becomes one-way, which exempts the communication time after the protocol. 

\section{Performance Evaluation}
\label{sec:evaluation}
We evaluate the performance of our protocols and explore the trade-off relationship between the parameters. The performance of our protocol depends on five parameters: the size of the 2D-array qubits $L$, the generation rate of entangled states $p_\mathrm{gen}$, the initial error rate of the physical entangled states $e_\mathrm{init}$, the error rate of the SWAP gates $e_\mathrm{swap}$, and the acceptance threshold $w_\mathrm{thr}$. The protocol is evaluated with two parameters: the success probability of the protocol $p_\mathrm{log}$, and the logical error rate of logical entangled states $e_\mathrm{log}$.

In this section, we first show the simulation settings and assumptions in our protocol. Then, we will show the performance of our protocols in several hardware configurations and explore available performance regions by tuning acceptance thresholds $w_\mathrm{thr}$.

\subsection{Simulation settings}
In this subsection, we explain the assumptions and settings used in our numerical evaluation. 
In the entanglement generation process, we assume that the entangled states are affected by uniform depolarizing noise. In other words, the entangled states are assumed to be generated with an initial error rate $e_\mathrm{init}$ and the initial states after entanglement generation are the Werner states with $1-e_\mathrm{init}$ of the fidelity.
\begin{equation}
    \hat{\rho}_\mathrm{init} = (1-e_\mathrm{init})\ket{\Phi^+}\bra{\Phi^+}+\frac{e_\mathrm{init}}{3}\sum_{i=1}^3 \sigma_A^{(i)}\ket{\Phi^+}\bra{\Phi^+} \sigma_A^{(i)}
\end{equation}
where $\ket{\Phi^+}=(\ket{00}+\ket{11})/\sqrt{2}$. Operators $(\sigma_A^{(1)}, \sigma_A^{(2)}, \sigma_A^{(3)})$ and $(\sigma_B^{(1)}, \sigma_B^{(2)}, \sigma_B^{(3)})$ denote Pauli operators $(X,Y,Z)$ on Alice's and Bob's qubit space, respectively. 

In the qubit rearrangement phase, we assume that each qubit of entangled pairs is affected by uniform depolarizing noise in each application of SWAP gates.
While Pauli errors might occur in Alice's and Bob's nodes independently, we can simplify the treatment of the noise model as follows. Since $\sigma^{(i)}_A\ket{\Phi^{+}}=\sigma^{(i)}_B\ket{\Phi^{+}}$, the error occurred on Alice's qubit can be rephrased as that on Bob's qubit. Thus, without loss of generality, we can assume only the SWAP gates in Bob's node suffer from the depolarizing noise with the following error rates.
\begin{equation}
    \tilde{e} = 2e_\mathrm{swap}-\frac{4}{3}e_\mathrm{swap}^2
    \label{eq:swap_error_rate}
\end{equation}

In the stabilizer measurement phase, we assume that the stabilizer measurements are noiseless for simplicity. In practice, this process would suffer from noisy stabilizer measurements, but we chose this assumption to simplify the numerical calculation.

The performance values, $p_{\rm log}$ and $e_{\rm log}$, are evaluated with Monte-Carlo sampling. We repeated $N=10^5$ trials to evaluate each parameter configuration. The numerical analysis procedure can be summarized as follows.
\begin{enumerate}
    \item We divide $N$ samples according to the chosen code distances. We denote the number of samples that constitute code distance $d$ as $N_d$.
    \item For each sample set with code distance $d$, we count the number of estimated errors to be no more than $w_\mathrm{thr}$, which we denote as $N_{\mathrm{acc}, d}$. The generating rate of a logical Bell state is calculated as $p_{\mathrm{log}, d}=N_{\mathrm{acc}, d}/N_d$.
    \item We estimate errors based on the obtained syndrome values and count the number of samples with logical errors $N_{\mathrm{err}, d}$. The logical error rate is calculated as $e_{\mathrm{log}, d}=N_{\mathrm{err}, d}/N_{\mathrm{acc}, d}$.
\end{enumerate}

\subsection{Numerical results}
Since there are many possible parameter combinations, we chose three hardware configurations consisting of the entanglement generation rate $p_{\rm gen}$, the lattice size $L$, or the initial infinity of the physical entanglement $e_{\rm init}$ to illustrate the performance of our protocol simply. The SWAP gate fidelity $e_{\rm swap}$ and the acceptance threshold $w_\mathrm{thr}$ are swept for each evaluation.

The performance depends on the chosen code distance, and showing all the plots for every sample set of code distances makes understanding difficult. To avoid this situation, the parameter set is chosen so that the code distance is typically concentrated to a certain one. In our numerical results, we only show the performance of the most frequent code distance for each hardware configuration.

\subsubsection{Configuration 1: $p_\mathrm{gen} = 0.3,\ L=19,\ e_\mathrm{init} = 0.05$ for code distance $d=7$}
We start the evaluation of the performance with parameters $p_\mathrm{gen} = 0.3,\ L=19,\ e_\mathrm{init} = 0.05$. 
We chose these values as a typical feasible value in the expected experimental setup (see Sec.\,\ref{sec:discussion} for experimental feasibility). Since most samples result in choosing $d=7$, we show the performance of the sample set of $d=7$.

We plotted the protocol success probability $p_\mathrm{log}$ and the logical error rate of the post-selected state $e_\mathrm{log}$ as a function of the error rate of SWAP gate $e_{\rm swap}$ in Fig.\,\ref{fig:experimental_results}~(a1) and (b1), respectively. Here, we varied $e_\mathrm{swap}$ from 0.05 to 0.005 and evaluated performances for several thresholds $w_\mathrm{thr}$.
As expected, as the acceptance threshold becomes large, i.e., allowing many estimated errors to occur, the post-selection probability increases while the logical fidelity is reduced. 
When we choose the acceptance threshold $w_\mathrm{thr} = 5$, we can prepare the logical entanglement while reducing the error rates from the physical ones, for example, if $e_\mathrm{swap}$ is below $0.03$, $e_{\rm log}\sim1.5\times10^{-2}$, and $p_{\rm log}\sim 0.11$.

Finally, we calculate the available tunability of the protocol success probability and the fidelity of logical entanglement by adjusting the acceptance threshold $w_\mathrm{thr}$. The results are plotted in Fig.\,\ref{fig:experimental_results}~(c1) for several $e_\mathrm{swap}$. 
This figure clearly illustrates the tunability of the generated rates and the quality of entanglement. The best choice of $w_\mathrm{thr}$ depends on the post-distillation process and the target logical error rates, which will be discussed in Sec.\,\ref{sec:post-process}.

\subsubsection{Configuration 2: $p_\mathrm{gen}=0.1,\ L=33, \ e_\mathrm{init}=0.05$ for code distance $d=7$}
Next, we chose the hardware configuration with $p_\mathrm{gen}=0.1,\ L=33, \ e_\mathrm{init}=0.05$. This configuration has the same initial error rates of physical entangled pairs and a similar average number of successfully entangled physical pairs $p_\mathrm{gen} L^2$, but has a larger lattice size $L$ and lower generation rates $p_\mathrm{gen}$ compared to Configuration 1. 

The results are shown in Figs.\,\ref{fig:experimental_results}~(a2), (b2), and (c2) in the same manner.
Note that the plots in Fig.\,\ref{fig:experimental_results} (a2) and (b2) for high error rate regions are not stable due to very low protocol success rates.
This configuration shows a lower protocol success probability and higher logical error rates, resulting in worse trade-off curves. This is because the successful qubit positions are scattered compared to Configuration 1, which means the required number of SWAP gates increases and degrades the results. 
Thus, we can conclude that a configuration with a small lattice size $L$ with high success probability $p_\mathrm{gen}$ is preferable to the opposite property in this parameter regime, while they achieve a similar average entanglement generation count.
It is also notable that even in this case, thanks to the post-selection mechanism of our protocol, the protocol can generate logical entangled states with higher fidelity than the initial fidelity if $e_{\rm swap}$ is sufficiently small.

\subsubsection{Configuration 3: $p_\mathrm{gen}=0.3,\ L=23, \ e_\mathrm{init}=0.05$ for code distance $d=9$}
\begin{figure*}[htbp]
\centering
\includegraphics[width = \linewidth]{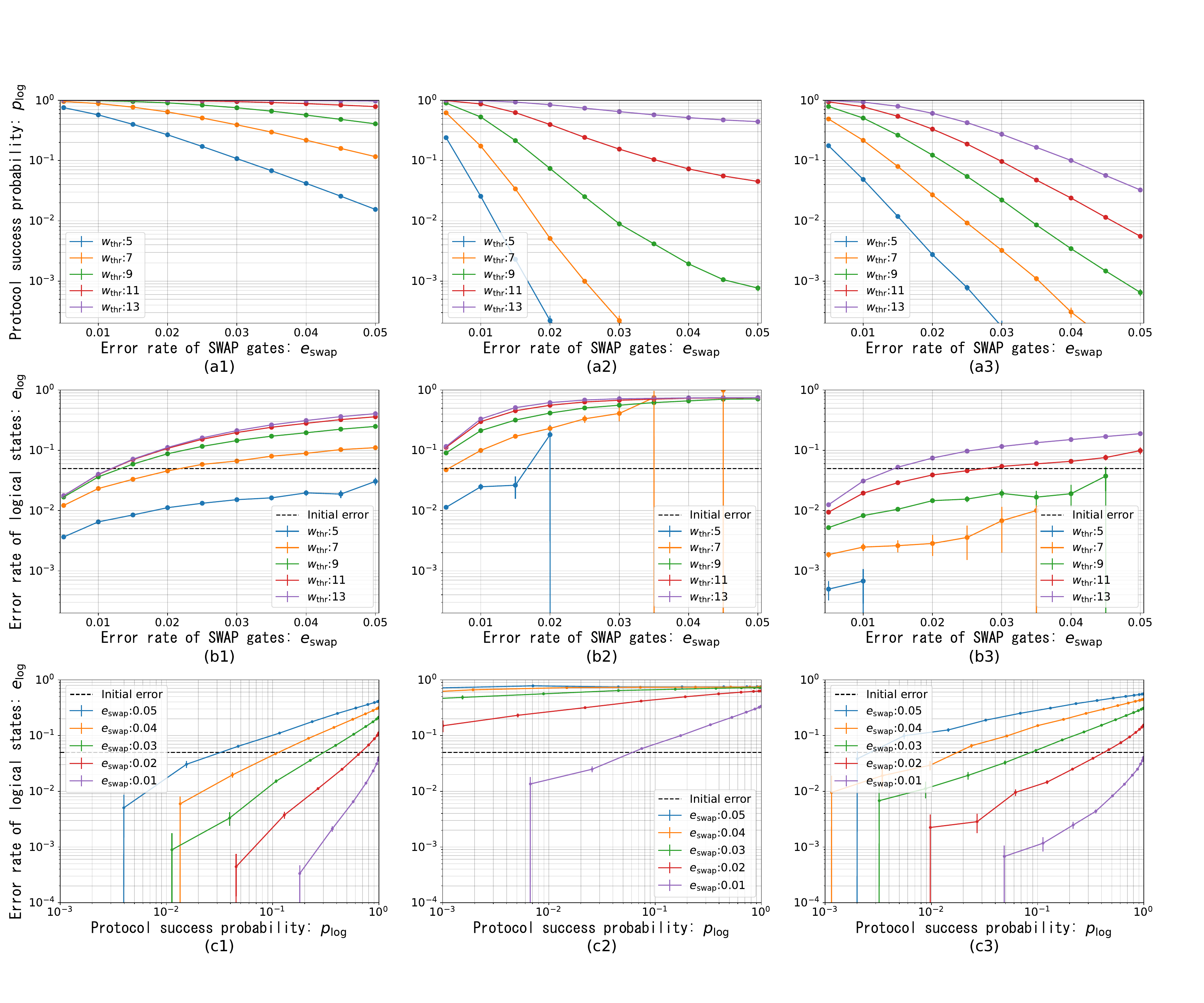}
\caption{(a1, a2, a3) show the process success probability $p_{\rm log}$ in terms of the error rate of SWAP gate $e_{\rm swap}$ for configurations 1, 2, and 3, respectively. (b1, b2, b3) show the logical error rate $e_{\rm log}$ for the error rates of SWAP gate $e_{\rm swap}$. (c1, c2, c3) show the trade-off relationships between $p_{\rm log}$ and $e_{\rm log}$ for configuration 1, 2, and 3, respectively.\\
The configurations 1, 2 and 3 represented by $(p_{\mathrm gen}, L, e_{\mathrm init}, d)$ are (0.3, 19, 0.05, 7), (0.1, 33, 0.05, 7) and (0.3, 23, 0.05, 9), respectively.}
\label{fig:experimental_results}
\end{figure*}

The final parameter set is $p_{\rm{gen}} = 0.3$, $L = 23$, and $e_{\rm{init}} = 0.05$, which typically constitutes $d = 9$ surface code. We chose this configuration to confirm the advantage of choosing a larger code distance in this protocol.
Fig.~\ref{fig:experimental_results}~(a3) and (b3) show the protocol success probability $p_{\rm log}$ and logical error rate $e_{\rm log}$ against the SWAP error rate $e_{\rm swap}$. The tunability between $p_{\rm log}$ and $e_{\rm log}$ is shown in Fig.~\ref{fig:experimental_results}~(c3).

Compared to Configuration 1, this configuration shows a lower logical error rate while the success probability is degraded. This is because the average error counts would increase as the number of data qubits in a surface-code cell increases. Thus, when we keep the acceptance threshold constant, an accepted event becomes rare, but we can ensure a smaller logical error rate as a code distance increases.
In the comparison of trade-off curves in Fig.~\ref{fig:experimental_results}~(a1) and (a3), we see that the available performance trade-offs in Configuration 1 are superior to those in Configuration 3. This indicates that large code distances do not necessarily provide better performance. We can expect the reason for this as follows. 
When physical error rates are smaller than the threshold value, surface codes with large code distances always provide better logical error rates without post-selection. On the other hand, physical error rates after physical entanglement sharing are typically higher than the threshold value in practice, as discussed in Sec.\,\ref{sec:discussion}. Thus, the advantage of using a large code distance is that it allows for stronger post-selection and small logical error rates that are not achievable in small code distances by sacrificing the success probabilities. However, there is a penalty for using large code distances since it demands more SWAP gates at the rearrangement step, which increases the effective error rate per entanglement at the stabilizer measurement step and can lose the advantage of a large code distance. This observation indicates that appropriate code distances should be chosen to explore the best trade-off relations even if there are a number of 2D channels.

\section{Analysis of communication rate}
\label{sec:discussion}

\subsection{Feasible parameter set in neutral atom systems}
\label{sec:communication_rate}
We discuss the experimental feasibility of our protocol in neutral atom systems. We suppose that physical entangled state sharing is performed with the following entanglement generation protocol: Alice and Bob excite atoms in their nodes, let the atom emit a photon entangled with the atomic state, and photons are coupled to optical fibers. The collected photons are measured in the Bell basis using optical circuit and photodetectors at the intermediate nodes. This protocol will generate physical entanglement of atoms in two distant nodes. When the communication length is long, we can convert the frequency of photons to improve the transmission rate of fibers at the cost of finite conversion efficiency.

Suppose we repeat the entanglement generation process for a time duration $\tau$ with the repetition rate $\gamma$. The error rate due to decoherence during the memory time for prepared entanglement can be calculated from $\tau$ and the lifetime of atoms. The success probability $p_{\rm gen}$ of obtaining an entangled pair between a pair of sites at Alice and Bob is given in the following equation:
\begin{equation}
    p_{\rm gen} = 1-\left(1-\frac{1}{2}10^{-\alpha l/10}\eta_{\rm ph}^2\eta_{\rm det}^2\eta_{\rm cov}^2\right)^{\gamma\tau}
    \label{repetition_rate}
\end{equation}
In this equation, the photon collection efficiency $\eta_{\rm ph}$ is the coupling efficiency of an entangled photon emitted from each atom into an optical fiber.
The photon detection efficiency by photo detector is reperesented by $\eta_{\rm det}$, which is the efficiency of Bell measurements at an intermediate station between two distant qubits.
The quantum frequency conversion efficiency is given by $\eta_{\rm cov}$, which is below unity if we utilize frequency conversion.
The factor of $10^{-\alpha l/10}$ represents the transmittance of the fiber, where $l$ is the communication distance and $\alpha$ is the attenuation rate. 

According to Young {\it et al.}~\cite{saffman_analysis}, the photon collection efficiency $\eta_{\rm ph}$ of a single atom system in free space is estimated to be $\eta_{\rm ph}\sim0.1$.
We can achieve a high detection efficiency by using superconducting nanowire single-photon detectors~(SNSPDs)~\cite{sspd1,sspd2}, which achieves about $\eta_{\rm det}\sim0.90$. Note that if we assume a typical Bell measurement setup based on linear optics, the success probability of Bell measurements is halved.
The achieved conversion rate $\eta_{\rm cov}$ from the resonant frequency of atomic transitions to the telecom frequency is approximately 0.57~\cite{QFC_weinfurter}.
The attenuation rate $\alpha$ at the telecom wavelength is typically $\alpha\sim0.21~{\rm db/km}$. 
Although most of these values are demonstrated in simplified systems, Hartung {\it et al.}~\cite{rempe_cavity_array} reported generating entangled photons from an atom-array system coupled to cavity modes while retaining the efficiency of atom-photon entanglement $\eta_{\rm ph}\eta_{\det}\sim0.3$ per attempt. Thus, we expect the combination of state-of-the-art technologies on neutral atom arrays can be demonstrated in the future.

Two scenarios should be separately considered to evaluate practical preparation time $\tau$ and generation rate $p_\mathrm{gen}$. The first one is the case of distributed quantum computation, i.e., neutral atom systems are located comparably close.  
In this case, fiber transmission loss can be negligible, and we do not need to utilize frequency conversion. Thus, we can assume $\eta_{\rm cov}=1$ and $e^{-\alpha l/10}\sim 1$. 
The repetition rate $\gamma$ is determined from the duration of the set of local operations, such as cooling, state preparation and excitation of the atoms. They are expected to be $350~{\rm \mu s}$, $3~{\rm \mu s}$, and $21~{\rm ns}$, respectively~\cite{Entangling_33km}. Thus, we expect the repetition rate would be about $\gamma\sim85\ {\rm kHz}$.
When we target $p_\mathrm{gen}=0.3$ with this repetition rate, the waiting time should be $\tau\sim1.0\times10^{-3}~{\rm s}$, which is shorter than the coherence time of atoms~\cite{lukin_quantum_processor}, and the protocol would generate high-fidelity entanglement.

The second scenario is long-distant quantum communication, i.e., neutral atom systems are placed a long distance apart. In this case, the repetition rate is upper-bounded by the communication distance $l$. According to the existing experimental results between $33~{\rm km}$ distant nodes~\cite{Entangling_33km}, the repetition rate is upper-bounded by $\gamma\sim6.3~{\rm kHz}$ due to the traveling time of light. The preparing time to achieve $p_\mathrm{gen}=0.3$ is also extended to $\tau\sim0.21~{\rm s}$, which would significantly increase the initial error rate of physical entangled pairs.

Several parameters are expected to be improved in the near future as various approaches have been investigated. For example, the photon collection efficiency has a room for improvement using cavity systems~\cite{rempe_nondestructive_Bellmeasure,Lukin_nanophotonicWG}. According to Young {\it et al.}~\cite{saffman_analysis}, the photon collection efficiency by using the cavity is estimated to be $\eta_{\rm ph}\sim0.48$ in a feasible experimental setup.
This significantly improves the preparation time $\tau\sim4.3\times10^{-5}~{\rm s}$ for short-distance communication and $\tau\sim9.2\times10^{-3}~{\rm s}$ for $33~{\rm km}$ communication. Further exploration and evaluation of future progress are left as future work.

\subsection{Performance with post-distillation}
\label{sec:post-process}
In this section, we evaluate whether logical entangled states can be practical enough for distributed quantum computation by combining our protocol with a post-distillation process. This evaluation provides a concrete estimate of the time required to reach a target logical fidelity (e.g., $10^{-13}$). Although we utilize a basic $[[n, 1, d]]$ code for distillation to maintain a clear and manageable demonstration, this stage might be further optimized through more efficient protocols or integrated error-correction frameworks, which is left as future work.

To demonstrate quantum computational supremacy with fault-tolerant quantum computing, logical error rates of all the logical operations must be smaller than the inverse of the number of logical operations, which is in the order of $10^{-13}$ according to Ref.\,\cite{target_logical_fidelity}, for example. 
On the other hand, the error rates of logical entangled states achieved in Sec.\,\ref{sec:evaluation} is much larger than this value.
Thus, in realistic cases, the code distance of logical entangled states generated by the proposed protocols is immediately expanded to the same distance used for locally storing logical qubits, and logical entangled states are further distilled to sufficiently small logical error rates with the post-distillation process. 
In this section, we calculate the rate of generating logical entangled states with sufficiently small logical error rates when our protocol with Configuration 1 is combined with entanglement distillation protocols at the logical level.

We assume that error detection with the $[[n,1,d]]$ QEC code is used in the post-distillation process utilizing $n$ pre-distilled entangled states. We suppose that the logical error rate after our pre-distillation protocol, $e_\mathrm{log}$, is sufficiently smaller than unity, and we assume that events with any syndrome flip in the post-distillation are rejected. Note that we can ignore errors in logical operations at the post-distillation stage since they are encoded with surface codes with sufficiently large code distances. Then, the error rate after post-distillation is roughly $e_\mathrm{log}^d$, and the success probability is about $(1-e_\mathrm{log})^n$.
In this configuration, the average trial count of the pre-distillation protocol per post-distillation trial is given by $n/p_{\rm log}$. The average trial count of the post-distillation, i.e., the inversed probability where we do not obtain any syndrome flip, is $1/(1-e_{\rm log})^n$.
Thus, we can estimate the average trial number of pre-distillation protocol $N_\mathrm{trial}$ until we obtain one logical entangled state through post-distillation as follows:
\begin{equation}
    N_\mathrm{trial} = \frac{n}{p_\mathrm{log}}\cdot\frac{1}{(1-e_\mathrm{log})^n}
    \label{equ:trial_number}
\end{equation}

\begin{figure}[htbp]
    \centering
    \includegraphics[width = 0.45\textwidth]{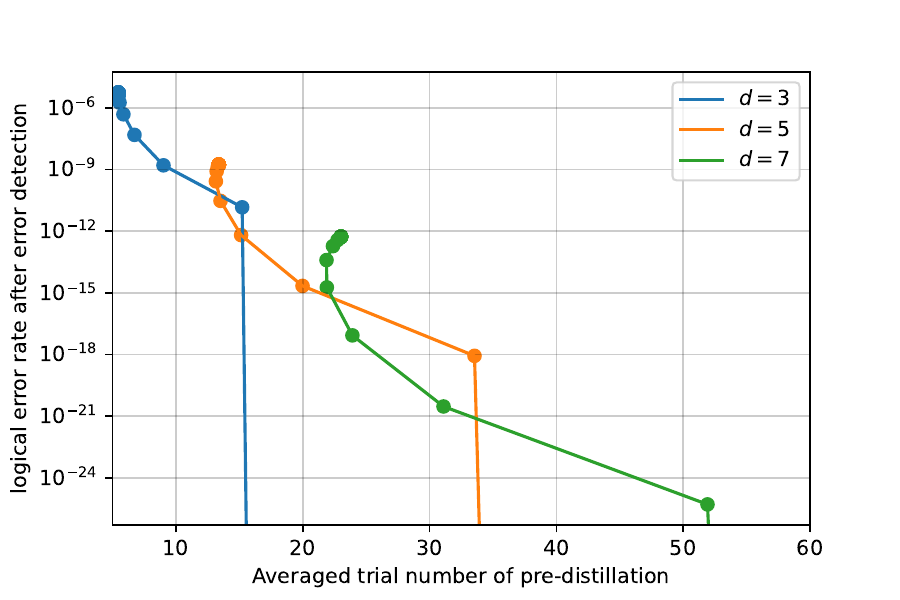}
    \caption{The achievable logical error rates are plotted as a function of averaged trial numbers of pre-distillation processes.}
    \label{fig:post_distillation}
\end{figure}

Combining the performance of Configuration 1, we can obtain the total trade-off performance as shown in Fig.~\ref{fig:post_distillation}.
The figure shows the trial number of pre-distillation $N_{\rm trial}$ to achieve a final target logical error rate with post-distillation under Configuration 1 with $e_{\rm swap}=5.0\times10^{-3}$. 
Blue, orange, and green lines correspond to the code distance of QEC codes for post-distillation protocols. Each point in the lines shows the performance with different acceptance thresholds $w_{\rm thr}$.
For post-distillation, we use QEC codes that realize code distance $d=3,5,7$ with minimal $n$, which is $n=5,11,17$, respectively~\cite{MAGMAtable,OriginalMAGMA}. We can see that the final logical error rates and total number of pre-distillations are in the trade-off relation. These results indicate that we should choose appropriate QEC codes for post-distillation according to the target logical error rates.

At several points in this figure, we can see that both logical error rates and trial numbers are simultaneously improved by changing the acceptance threshold $w_{\rm thr}$. Though this might seem inconsistent with the behavior of pre-distillation, this can be explained as follows.
When we decrease the logical error rate at the cost of increased trial counts of pre-distillation per post-distillation trial, the average trial count of post-distillation is improved. 
Since the total trial count of pre-distillation $N_{\rm trial}$ is determined as the product of the trial count of pre-distillation per post-distillation $N_0$ and that of post-distillation $N_1$, the total trial count $N_{\rm trial}$ can be sometimes improved when the reduction of $N_1$ overwhelms the increase of $N_0$.

\subsection{Logical communication bandwidth in neutral atom systems}
Combining the results of Sec.\,\ref{sec:communication_rate} and Sec.\,\ref{sec:post-process}, we can calculate a quantitative communication bandwidth. Here, we calculate the bandwidth when two nodes are close.
According to Fig.\,\ref{fig:post_distillation}, we can achieve logical error rate $2.2\times10^{-15}$ with average trial count $N_{\rm trial}\sim20$ with $d=5$. 
Each pre-distillation trial consists of the time for entanglement preparation $\tau$, qubit arrangement $\tau_{\rm arr}$, and measurement $\tau_{\rm meas}$. 
The preparation time $\tau$ of logical entangled states for a short distance is estimated as $1.0 \times 10^{-3}\,{\rm s}$ as discussed in Sec.~\ref{sec:communication_rate}. 
According to our simulation, the qubit arrangement time $\tau_{\rm arr}$ typically needs a few hundred SWAP gates, which we estimated as $\tau_{\rm arr} \sim 1\,{\rm ms}$. We also estimated $\tau_{\rm meas} \sim 1\,{\rm ms}$. 
In this evaluation, we neglected the latency for stabilizer measurements on surface-code logical qubits in the post-distillation process since they can be performed with several transversal CNOTs to ancillary logical qubits and single simultaneous measurements on them, which is expected to be much shorter than the time for pre-distillation per single post-distillation trial.
From these values, the average preparation time of post-distilled logical entangled states with fidelity $2.2\times10^{-15}$ is estimated as $60~{\rm ms}$, corresponding to the generation rate $17~{\rm Hz}$. 

If we can assume the assistance of cavity, the preparation time $\tau$ and measurement $\tau_{\rm meas}$ can be shorter.
Suppose that $\tau$ is reduced to $4.3 \times 10^{-5}\,{\rm s}$ as given in Sec.~\ref{sec:communication_rate}, and $\tau_{\rm meas}$ becomes $100\,{\rm \mu s}$~\cite{cavity_based_measurement_Rempe}. Then, the rate of logical entangled states is improved to $44~{\rm Hz}$.

It is worth noting that the communication rate is affected by both collection rates of photons and latencies of local operations. If two nodes are close and local operations are time-consuming, a fast entanglement generation rate by a high photon collection rate does not drastically improve communication speed.

\section{Related work}
\label{sec:related_work}
\begin{figure*}[htbp]
    \centering
    \includegraphics[width = 1.0\linewidth]{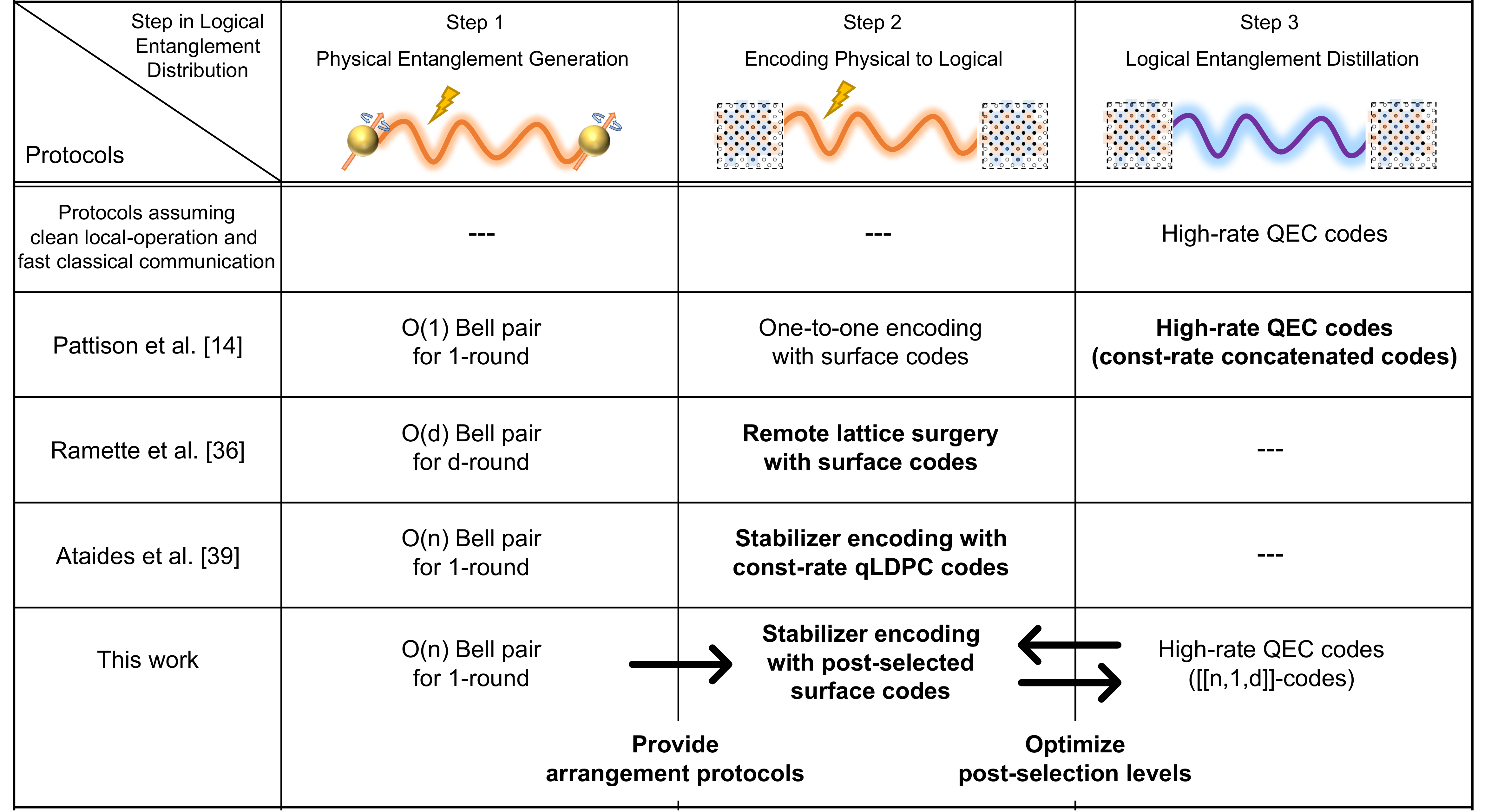}
    \caption{The focus of existing papers and the positioning of our work. Focused points of each paper are written in bold font.}
    \label{fig:positioning}
\end{figure*}

Several proposals have been made to design efficient protocols for distributing logical entanglement. As explained in Sec.\,\ref{sec:introduction}, the overall logical entanglement distribution protocol can be divided into three steps: (1) distribution of noisy physical Bell pairs, (2) encoding of noisy physical Bell pairs into noisy logical Bell pairs, and (3) distillation of noisy logical Bell pairs into clean logical Bell pairs. The ultimate goal in designing this protocol is to maximize the generation rate of logical Bell pairs with a target fidelity. This paper focuses on Step~2 and proposes fast and tunable protocols to offer more flexibility in global optimization.
While both Steps~2 and 3 require QEC codes for encoding and distillation, their objectives differ. QEC codes in Step~2 must account for local errors as well as protocol repetition and encoding rates. In contrast, Step~3 can neglect local errors because the QEC codes used in Step~2 already protect qubits. Thus, Step~3 primarily requires high encoding rates. Consequently, the two steps demand distinct optimization strategies. This section reviews representative schemes proposed in the literature on FTQC. See Fig.\,\ref{fig:positioning} for the positioning of our work.

Most existing studies on logical entanglement distribution assume that local operations are error-free and aim to maximize encoding rates, i.e., the ratio between the number of clean output entangled pairs and that of noisy input pairs. This situation is justified when we focus on Step~3 and assume that the input noisy logical entanglement is generated much more slowly than internal logical operations. It is well known that finding distillation protocols with a high encoding rate is equivalent to finding QEC codes with a high encoding rate, and general theoretical frameworks have been extensively discussed~\cite{purification_bennett,Quantum_repeaters_based_on_purification,stabilizer_distillation,knill2005scalable}. As theories of constant-rate QEC codes have been developed recently~\cite{yamasaki2024time}, several constant-rate distillation protocols have been proposed. 
Refs.\,\cite{pattison2025constant,gu2025constant} present constant-rate constructions using concatenated codes and scrambling techniques. However, these works typically do not consider the implementation of Step~2, or employ only simple strategies for it. For example, Ref.\,\cite{pattison2025constant} adopts one-by-one encoding~\cite{li2015magic} for Step~2, which results in fidelities comparable to those of raw physical Bell pairs. Table~I in Ref.\,\cite{pattison2025constant} shows that the distillation input error rate (i.e., the logical error rate after Step~2) strongly impacts the performance of Step~3. Hence, existing studies suggest that cross-step optimization is crucial for improving total performance. Since our work expands the design space of Step~2, our protocol can be integrated with these schemes to enhance performance and tunability, rather than conflicting with them.

Refs.\,\cite{fault-tolerant_connection,leone2025resource,fault_tolerant_optical_interconnects} investigate methods based on remote lattice surgery, which performs lattice surgery~\cite{horsman2012surface} between two distant code blocks by linking them through physical remote CNOTs consuming physical Bell pairs. The advantage of remote lattice surgery is that high-fidelity logical Bell pairs can be generated directly from physical Bell pairs, provided that the error rate of physical Bell pairs is below the threshold and sufficiently large code distances are available. A drawback of this approach is that it requires a large number of Bell pairs if we assume practical physical error rates~\cite{pattison2025constant}. While our protocol focuses on encoding via stabilizer measurements with post-selection, it may be beneficial to use remote lattice surgery with a small code distance as an encoding phase, followed by distillation in Step~3 to reach the target logical error rate.

Recently, Ref.\,\cite{ataides2025constant} independently proposed a protocol addressing Step~2. This work employs constant-rate quantum low-density parity-check (qLDPC) codes and performs syndrome projections to map physical Bell pairs onto logical Bell pairs~\cite{stabilizer_distillation}. Although this approach achieves a high encoding rate in Step~2, it faces challenges inherent to qLDPC codes. 
The protocol requires a fault-tolerant implementation of the chosen qLDPC codes. While progress has been made toward such implementations, their integration into a full entanglement distribution network remains a significant challenge. This is primarily because qLDPC codes typically necessitate high-connectivity hardware, such as moving optical tweezers or long-range interconnects, to handle their non-local stabilizer measurements.
More importantly, since multiple logical Bell pairs are embedded within a single large code block in Step 2 of Ref.\,\cite{ataides2025constant}, performing subsequent operations, such as entanglement distillation in Step 3, becomes highly nontrivial. Such a structure complicates cross-step optimization, as operations on individual logical pairs may require complex routing or joint decoding within the same block.
In contrast, our protocol utilizes a more modular approach based on surface codes, where logical states are handled as independent units. This allows for straightforward optimization of the end-to-end performance using experimentally feasible components and parameters. Thus, Ref.\,\cite{ataides2025constant} complements our work: our approach is more feasible for near-term implementation in 2D architectures, whereas theirs represents a high-performance alternative for platforms providing the necessary connectivity to support integrated qLDPC operations in the future.

In summary, the main contributions of our paper are: (1) proposing a concrete Step~2 protocol that is tunable and implementable on 2D devices, (2) performing cross-step optimization across the logical-entanglement-distribution stages, and (3) numerically evaluating practical performance after optimization based on experimentally feasible components and parameters. Although the quantitative results reported here may be further improved by future advances, such as efficient qLDPC codes, enhanced post-selection, and improved cavity-QED interfaces, our protocol provides a fundamental methodology for evaluating global optimization in logical entanglement distillation, contributing to the future design of distributed quantum computers.

\section{Conclusion}
\label{sec:conclusion}
We proposed a logical entanglement distribution protocol tailored to 2D integrated qubits, which includes rearranging distributed entangled states by SWAP gate operations to encode surface code and post-selection for distilling logical entangled states. Our protocol offers the tunability between the quality and speed of logical entanglement sharing, which can be used for optimizing the post-distillation process.
We numerically examined the performance of our protocol by targeting neutral atom systems, and also evaluated the performance of the combination of our protocol and post-distillation processes.
As a result, we showed that the protocol effectively addresses the challenges posed by probabilistic entanglement generation and the nearest-neighboring two-qubit quantum operation. The tunability of our protocol between protocol success probability and fidelity will contribute to the feasibility of near-future experimental setups. 

Our protocol can be naturally applied to other qubit devices with a 2D array of physical communication channels. 
For example, for short-distance communication, entanglement generation between two superconducting qubits in remote cryostats has been demonstrated~\cite{old_entangling_superconducting_qubits_Nature,entangling_superconductor_qubits}. 
According to Magnard {\it et al.}~\cite{entangling_superconductor_qubits}, the error rate of the entangled state, efficiency of entangling attempts, and repetition rate are $e=0.205,~\eta=0.675,~\gamma=6.2$~MHz, respectively. 
The high-performance design of logical communication protocol under this parameter region is left as future work.


We evaluated the performance of our protocol considering essential sources of physical error, such as local SWAP errors and the infidelity of physical Bell pairs. However, real quantum devices suffer from additional types of errors, including local circuit errors in stabilizer measurements. 
Our results show that this protocol is tolerant to errors in entanglement generations and movements for further operations.
Although we believe that the current noise models qualitatively capture the essential behavior of our protocol, it would be desirable to evaluate it under more realistic noise models. Nevertheless, estimating logical error rates under such complex models is computationally challenging, because the number of local SWAP operations varies from one sample to another, making it difficult to employ standard software libraries such as Stim~\cite{gidney2021stim} to speed up the evaluation. Developing efficient methods for evaluating performance under these complex noise models is an important direction for future work.

While this paper focused on local SWAP gates driven by optical pulses, an alternative rearrangement protocol can be designed based on optical tweezers~\cite{lukin_quantum_processor} when neutral atoms are assumed. We did not assume shuttling technologies in this paper because they are specific to certain platforms, such as neutral atoms, and the reduction of logical error rates via shuttling has not yet been experimentally demonstrated. Moreover, the use of moving tweezers is not yet widely established in experimental practice~\cite{saffman2025quantum}. Nevertheless, exploring protocols that assume high-fidelity and fast shuttling technologies could serve as effective device-specific optimizations. 
It is not straightforward to determine whether the use of moving tweezers can improve our protocol, since there is no intrinsic parallelism in the rearrangement phase and shuttling is slower than optically driven SWAP gates. Although several rearrangement protocols for forming fully occupied qubit arrays have been proposed~\cite{lukin_quantum_processor,48logical_qubits_Lukin,barredo2016atom,lin2025ai}, our problem setting differs from these studies because we must move entangled qubits to data-qubit positions and other qubits to ancillary-qubit positions. Therefore, exploring more efficient rearrangement schemes remains an interesting direction for future research.

Another promising direction for improving overall performance is the use of qLDPC codes. Fault-tolerant implementation of qLDPC codes typically relies on advanced technologies such as fast and reliable optical tweezers~\cite{atom_qldpc} or long-range interconnects between superconducting qubits~\cite{high-threshold_and_low-overhead}, which must achieve physical error rates below threshold. As shown in Ref.~\cite{ataides2025constant}, qLDPC codes can provide higher encoding rates but require advanced technologies and additional challenges in the post-distillation process (i.e., Step~3 in Fig.~\ref{fig:sketch_of_distribution}), since directly available logical operations on qLDPC codes are limited. Evaluating the total performance while accounting for these restricted logical operations is also an interesting topic for future investigation.

While our post-selection criterion is based on the estimated error count, other metrics are also possible. For instance, one may use the number of flipped syndrome values, which is simpler to compute than the estimated error count and improves the repetition rates if the time for error estimation is a bottleneck. Another candidate is the complementary gap~\cite{bombin2024fault,gidney2025yoked}, defined as the number of additional physical errors required to alter the outcome of a minimum-distance decoding algorithm. Although calculating the complementary gap introduces additional latency, it accurately reflects the reliability of the encoding process. As long as QEC codes can be fault-tolerantly implemented on real devices, our post-selection mechanism and these alternative criteria can be universally defined for any stabilizer code~\cite{stabilizer_distillation,purification_bennett}, providing a general framework to enhance the reliability of logical-state encoding.

\section*{Acknowledgement}
This work was supported by PRESTO JST Grant No. JPMJPR1916, MEXT Q-LEAP Grant No. JPMXS0120319794 and JPMXS0118068682, JST Moonshot R \& D Grant No. JPMJMS2061 and JPMJMS2066, JST CREST Grant No. JPMJCR23I4 and JPMJCR24I4 and Program for Leading Graduate Schools Interactive Materials Science Cadet Program.

\section*{Date Avalability}
The source code for the numerical simulations and the scripts used to generate the figures in this study are available on GitHub \url{https://github.com/Myuya-academia/logical-2d}.

\bibliography{reference.bib}

\end{document}